**Self-organized clustering, prediction, and superposition of long-term cognitive decline from short-term individual cognitive test scores in Alzheimer's disease**

Hiroyuki Sato[1], Keisuke Suzuki[2], Atsushi Hashizume[3,4], Ryoichi Hanazawa[1], Masanao Sasaki[1], Akihiro Hirakawa[1,#], the Japanese Alzheimer's Disease Neuroimaging Initiative*, and the Alzheimer's Disease Neuroimaging Initiative**

[1]Department of Clinical Biostatistics, Graduate School of Medical and Dental Science, Tokyo Medical and Dental University, Bunkyo-ku, Tokyo, Japan
[2]Innovation Center for Translational Research, National Center for Geriatrics and Gerontology, Obu, Aichi, Japan
[3]Department of Clinical Research Education, Nagoya University Graduate School of Medicine, Nagoya, Aichi, Japan
[4]Department of Neurology, Nagoya University Graduate School of Medicine, Nagoya, Aichi, Japan

#Corresponding author
Akihiro Hirakawa, Department of Clinical Biostatistics, Graduate School of Medical and Dental Science, Tokyo Medical and Dental University, 1-5-45 Yushima, Bunkyo-ku, Tokyo 113-8510, Japan, Phone: +81-3-5803-5150, Fax: +81-3-5803-0421, E-mail: a-hirakawa.crc@tmd.ac.jp

*The full membership of the Japanese Alzheimer's Disease Neuroimaging Initiative (ADNI) investigators is listed at: https://humandbs.biosciencedbc.jp/en/hum0043-j-adni-authors.
** Data used in preparation of this article were obtained from the Alzheimer's Disease Neuroimaging Initiative (ADNI) database (adni.loni.usc.edu). As such, the investigators within the ADNI contributed to the design and implementation of ADNI and/or provided data but did not participate in analysis or writing of this report. A complete listing of ADNI investigators can be found at: http://adni.loni.usc.edu/wp-content/uploads/how_to_apply/ADNI_Acknowledgement_List.pdf

**Abstract**
Progressive cognitive decline spanning across decades is characteristic of Alzheimer's disease (AD). Various predictive models have been designed to realize its early onset and study the long-term trajectories of cognitive test scores across populations of interest. Research efforts have been geared towards superimposing patients' cognitive test scores with the long-term trajectory denoting gradual cognitive decline, while considering the heterogeneity of AD. Multiple trajectories representing cognitive assessment for the long-term have been developed based on various parameters, highlighting the importance of classifying several groups based on disease progression patterns. In this study, a novel method capable of self-organized prediction, classification, and the overlay of long-term cognitive trajectories based on short-term individual data was developed, based on statistical and differential equation modeling. We validated the predictive accuracy of the proposed method for the long-term trajectory of cognitive test score results on two cohorts: the Alzheimer's Disease Neuroimaging Initiative (ADNI) study and the Japanese ADNI study. We also presented two practical illustrations of the simultaneous evaluation of risk factor associated with both the onset and the longitudinal progression of AD, and an innovative randomized controlled trial design for AD that standardizes the heterogeneity of patients enrolled in a clinical trial. These resources would improve the power of statistical hypothesis testing and help evaluate the therapeutic effect. The application of predicting the trajectory of longitudinal disease progression goes beyond AD, and is especially relevant for progressive and neurodegenerative disorders.

**Introduction**

The neurodegenerative symptoms of Alzheimer's disease (AD) are characterized by a progressive cognitive decline spanning several decades, underscoring the importance of research into its early detection and essential therapeutic or preventive interventions. However, the practicality of conducting periodic, long-term follow-ups for cognitively healthy individuals over 20–30 years is a significant challenge, leading to the need for developing a prediction methodology or framework for assessing cognitive decline based on short-term follow-up data on cognitive function.

Recent prediction studies on AD can be categorized based on their objectives and the data used. A first such category would be to predict an onset of AD based on clinical and neuroimaging data over a primarily three-year assessment period of cognitive impairments[1]. The second is the study on predicting the cognitive decline of cognitive test score within 2–8 years using clinical, biomarker, and imaging data in mild cognitive impairment (MCI) patients[2–4]. The third, as focused in this study, aimed to predict the long-term temporal course (such as over 10 years) of cognitive test scores of interest spanning from cognitively normal to AD[5–8]. These studies had sought to address the challenge of predicting long-term trajectories of cognitive test scores based on short-term individual data (e.g., 2–3 years), using nonlinear mixed-effects models. However, these nonlinear mixed-effects models often yield varied in parameter estimates (i.e., the estimated trajectory) depending on the initial values provided during parameter estimation. The practical effectiveness of these methods in contributing to the early detection of AD or the development of therapeutic treatments for AD has not been sufficiently demonstrated. In addition to these approaches, a novel statistical method for predicting long-term trajectories based on a differential equation method has been recently developed[9]. This method can stably estimate the longitudinal trajectory even in a small sample size, although the successful estimation using nonlinear mixed-effects models depends on the initial values given in the parameter estimation.

The awareness of heterogeneity within AD underscores the imperative to classify the long-term trajectory of cognitive test scores into distinct types with varying trajectories. Numerous studies have explored AD-specific signatures by clustering cross-sectional data from neuropathological[10], neuroimaging[11,12], clinical[13], and biochemical[14] biomarkers. Recently, Poulakis et al.[15] described five distinct longitudinal patterns of brain atrophy associated with distinct demographic and cognitive profiles, shedding light on the existence of differential stages and subtypes of AD. However, to develop a prediction method to forecast the long-term trajectory of cognitive test scores while accounting for the disease heterogeneity, classifying the long-term trajectory into several groups with different disease progression patterns becomes necessary, but the existing methods[5-9] fall short of achieving this. Moreover, beyond prediction and classification, there is a need to identify the group of trajectories to which individual patients belong, and the extent to which they align with the projected trajectories. Superimposing the observed cognitive test scores on individuals and the corresponding predicted trajectory can provide information on how cognitive function declines in each patient.

To fulfill the three key requirements for an optimal model—including prediction, classification, and superposition of long-term cognitive trajectory from short-term follow-up data—we developed a novel methodology. This approach demonstrates self-organized prediction, classification, and superposition of long-term cognitive trajectories based on short-term individual data. By leveraging statistical and differential equation models, and further, extending the recently-proposed method[9], its four steps can help determine the multiple latent long-term trajectories of these scores based on only short-term data, employing a self-organized approach. This method would also allow us to estimate which specific point in the trajectory belong to each individual. For instance, with baseline, 6-month, and 1-year follow-up data on cognitive test scores for individuals at risk for AD, the proposed method would enable the predicting of which trajectory was followed, and at what point they currently are at, based on the rate of cognitive decline over a year. We validate the proposed method using the Clinical Dementia Rating Sum of Boxes (CDR-SOB), Mini-Mental State Examination (MMSE), and 13-item Alzheimer's Disease Assessment Scale-Cognitive (ADAS-Cog) in two cohorts with different sample sizes and follow-up periods: the Alzheimer's Disease Neuroimaging Initiative study (ADNI) and Japanese ADNI (J-ADNI). Additionally, we present two practical illustrations using the proposed method: (i) the simultaneous evaluation of apolipoprotein E (APOE) ε4 status associated with both the onset and longitudinal progression of AD, and (ii) the development of an innovative randomized controlled trial (RCT) design aimed at mitigating the heterogeneity in disease progression among enrolled patients, thereby improving the power of statistical hypothesis testing for evaluating therapeutic effects. We propose that the prediction-classification-superposition of cognitive decline would contribute to the early detection of cognitive decline and the evaluation of therapeutic interventions for AD.

**Results:**

**Population analyzed**

We extracted participants with the available data on baseline assessments of diagnosis, age, sex, length of education, APOE ε4 status, and β-amyloid peptide (Aβ) status from the ADNI and J-ADNI cohorts. Our sample included 1,503 individuals (336 normal cognitive (NC), 176 subjective memory complaints (SMC), 337 Early MCI (EMCI), 385 Late MCI (LMCI), and 269 AD) in the ADNI group, and 288 individuals (84 NC, 116 MCI, and 88 AD) in the J-ADNI group. Participant demographics in the ADNI and J-ADNI cohorts were similar, except for the proportion of participants with Aβ-positive MCI and AD (Tables 1 and 2). The participants' demographic characteristics by Aβ and APOE ε4 status are shown in Tables 3 and 4. For each cognitive test score, we defined participants with Aβ positivity, having both baseline and at least two post-baseline measurements. The analysis set for estimating the long-term trajectory is presented in Table 5 and Figure 1.

**Four steps of the proposed method**

**Step I: Clustering of decline rate of cognitive test score in individuals**

We estimated the individual decline rate of cognitive test scores (i.e., MMSE, ADAS-Cog, and CDR-SOB in the ADNI and J-ADNI cohorts) based on the mixed-effects model for the change in score from the baseline value (see "Statistical descriptions of the four steps" section in the Methods chapter). This model includes the interaction term of the disease stage of the participants (i.e., NC, SMC, EMCI, LMCI, and AD for the ADNI; NC, MCI, and AD for the J-ADNI) and the (continuous) follow-up time as the fixed (i.e., group-specific) effect; further, the slope of cognitive decline denotes the random (i.e., individual-specific) effects. For each participant, the sum of fixed and random effects ($\tilde{\beta}_i = \mathbf{x}_i^T \hat{\boldsymbol{\beta}} + \hat{b}_i$ in the "Statistical descriptions of the four steps" section) is the estimate of slopes of cognitive decline (i.e., decline rate) based on the cognitive test score. In this proposed method, these estimated slopes of cognitive decline in individuals are categorized into several groups according to the degree of cognitive decline for classification. For simplicity, but not limited to it, we categorized participants into moderate, intermediate, and rapid cognitive decline groups, respectively for each cognitive test score, as follows: −2 to 0, −4 to −2, and < −4 for the MMSE; 0 to 3, 3 to 6, and 6 ≤ for the ADAS-Cog; and 0 to 1, 1 to 2, and 2 ≤ for the CDR-SOB. For each cognitive decline group, the relationship between the estimated slope of cognitive decline and the average cognitive test score during the follow-up period was analyzed in Step II (Figure 2 and Table 6).

**Step II: Modeling relationship between cognitive decline rate and average scores**

To estimate the long-term trajectories of cognitive test scores in the three cognitive decline groups categorized in the previous step, we modeled the relationship between the slope of cognitive decline and the average cognitive test scores during the follow-up period. The assumed statistical model for this relationship determines the form of the longitudinal trajectory of the cognitive test scores. In nature, the cognitive function gradually decreases and then reaches a plateau; therefore, the slope of cognitive decline was assumed to be zero for the most affected cognitive test scores (0 for MMSE, 85 for ADAS-Cog, and 18 for CDR-SOB). Thus, in terms of modeling MMSE, for each participant, we assumed the statistical model of [slope = group-specific coefficients × {(30 − average MMSE) × average MMSE / 30} + random error] (Equation (2) from the "Statistical descriptions of the four steps" section), where group-specific coefficients are random effect parameters; similarly, in terms of modeling ADAS-Cog and CDR-SOB, we assumed the statistical model of [slope = group-specific coefficients × {(average ADAS-Cog − 85) × average ADAS-Cog / 85} + random error] and [slope = group-specific coefficients × {(average CDR-SOB − 18) × average CDR-SOB / 18} + random error], respectively (Equations (3) and (4) from the "Statistical descriptions of the four steps" section). The mean curves estimated from these statistical models for the moderate, intermediate, and rapid groups are respectively shown as black, blue, and red solid lines in Figure 2. Notably, the different statistical models characterize the different forms of the longitudinal trajectory; therefore, we can select the appropriate model based on the criteria for quantifying the goodness of the trajectory fit in practice (introduced in Step VI).

**Step III: Prediction of long-term trajectory of cognitive test scores**

The predicted cognitive test score corresponding to its disease progression time (i.e., years from disease onset) was estimated by solving for the ordinary differential equation including the parameters for group-specific coefficients estimated in the Step II (i.e., solving for the differential equation of $d\mu(\tau)/d\tau = \phi(\hat{\mu}_i; \hat{\gamma}_s)$ shown in the "Statistical descriptions of the four steps" section). Using the paired data of the predicted cognitive test scores and the corresponding disease progression time, we depicted the longitudinal trajectory of the cognitive test scores between the specific intervals (Figure 3).

**Step IV: Superposition of the observed cognitive test scores in individuals on long-term cognitive decline trajectory**

Finally, we superposed the observed cognitive test scores of the individuals on the predicted longitudinal trajectory. For each participant in each cognitive decline group, we selected the predicted cognitive test score on the trajectory that was closest to the observed score at baseline, and then superposed the post-baseline observed data on the corresponding predicted longitudinal trajectory (Figure 3). Furthermore, we quantified the goodness of trajectory fit using the average predictive error (APE) that is the square root of mean square error of difference between the estimated and observed scores in all observation timepoints across the participants ($APE = \sqrt{\frac{1}{n_s}\sum_{i=1}^{n_s}\left[\frac{1}{J_i}\sum_{j=0}^{J_i}\{y_{ij} - \hat{\mu}_s(\tau_{ij})\}^2\right]}$ in the "Statistical descriptions of the four steps" section).

**Prediction accuracy**

To examine the prediction accuracy of the proposed method, we conducted the internal and external validations of our prediction models. For internal validation, we first randomly split the ADNI (or J-ADNI) data into five equally-sized folds of cross validation. We reserved one of the folds as a test set and combined the remaining four folds into a training set. We developed the prediction model based on the proposed method using the training set and calculated the APEs on the test set. We repeated these steps 1,000 times and calculated the average APEs across 1,000 random five-fold cross validations. For external validation, we also calculated the APE of the prediction models developed from the ADNI (or J-ADNI) data by applying them respectively to the J-ADNI (or ADNI) data. Table 7 shows the results of the internal and external validations of the prediction models for the MMSE, ADAS-Cog, and CDR-SOB. Internal validations for the ADNI cohort showed the average APEs to be 2.12 to 3.64 points for MMSE, 3.82 to 5.83 points for ADAS-Cog, and 1.06 to 1.99 points for CDR-SOB, respectively. As for the external validation for the ADNI cohort using J-ADNI data, the APEs were 1.62 to 2.48 points for MMSE, 2.69 to 3.81 points for ADAS-Cog, and 0.67 to 2.34 points for CDR-SOB, respectively. For all cognitive test scores, the APEs increased with an increasing slope of cognitive decline, but all cognitive decline groups could control the APEs within almost 10% of the maximum of each score (3 points for MMSE, 8.5 points for ADAS-Cog, and 1.8 points for CDR-SOB). Similar tendencies were observed in the internal and external validation for the J-ADNI cohort. As shown in Table 6 and Figure 2, several patients with rapid cognitive decline showed a much larger slope than the group average of the estimated slope, which might have led to the larger APE.

**Application 1: Simultaneous evaluation of association between APOE ε4 status and the onset/ longitudinal progression of AD**

We illustrated the application of the proposed method to simultaneously evaluate the association between a patient background factor and the onset/longitudinal progression of AD in the ADNI participants with Aβ-positive EMCI and LMCI, as shown in Table 3, although the factors associated with onset[16–19] and longitudinal progression[20–22] of AD are often assessed in separate analyses. For example of using APOE ε4 status, we estimated the longitudinal trajectories of the MMSE, ADAS-Cog, and CDR-SOB scores based on the status of APOE ε4 in participants with Aβ-positive EMCI and LMCI of the ADNI cohort (Figure 4). Based on these estimated longitudinal trajectories in participants with Aβ-positive MCI (i.e., EMCI and LMCI), we could simultaneously compare the two populations with and without APOE ε4 in terms of (i) the time to reach a milestone for possible AD onset (23 points for MMSE, 30 points for ADAS-Cog, and 4.5 points for CDR-SOB) and (ii) the area under the long-term trajectory curve (AUC) between zero year and a specified time point in a year, which quantifies the amount of longitudinal progression of AD. The ratio of AUCs between the Aβ-positive MCI population with and without APOE ε4 indicates the relative risk of disease progression during that period.

For participants with Aβ-positive MCI of the ADNI cohort, the time to reach a milestone for possible AD onset, along with 95% confidence intervals (CIs) and two-sided p-values estimated by the bootstrap method, are shown in Table 8. The time to reach 23 points of MMSE in the Aβ-positive MCI population with and without APOE ε4 were 6.4 (95%CI, 5.6 to 7.4) and 6.3 (95%CI, 5.1 to 8.4) years, respectively. For CDR-SOB, the time to reach 4.5 points in the Aβ-positive MCI population with and without APOE ε4 were 6.3 (95%CI, 5.8 to 7.1) and 6.9 (95%CI, 5.8 to 9.3) years, respectively. There were no significant differences in the time to reach a milestone for a possible AD onset between the Aβ-positive MCI population with and without APOE ε4 based on MMSE and CDR-SOB (difference: 0.1 years, two-sided p-value = 0.836 for MMSE, difference: –0.6 years, two-sided p-value = 0.222 for CDR-SOB). In contrast, based on ADAS-Cog, the time to reach 30 points was significantly earlier in the Aβ-positive MCI population with APOE ε4 than in the Aβ-positive MCI population without APOE ε4: 17.1 (95%CI, 14.9 to 20.0) years in the Aβ-positive MCI population with APOE ε4 and 23.6 (95%CI, 17.5 to 35.0) years in the Aβ-positive MCI population without APOE ε4 (difference: −6.5 years, two-sided p-value = 0.012). For the longitudinal trajectories of each cognitive test in the Aβ-positive MCI population of the ADNI cohort shown in Figure 4, the ratio of AUCs for ADAS-Cog in the Aβ-positive MCI population with APOE ε4 relative to that without APOE ε4 demonstrated statistical significance (i.e., AUC ratio ≠ 1), but not for MMSE and CDR-SOB (AUC ratio: 1.34, two-sided p-value = 0.014 for ADAS-Cog, AUC ratio: 1.01, two-sided p-value = 0.798 for MMSE, AUC ratio: 1.03, two-sided p-value = 0.178 for CDR-SOB).

We also evaluated the association between APOE ε4 status and longitudinal progression of AD in the Aβ-positive AD population of the ADNI cohort. The longitudinal trajectories of each cognitive test for the Aβ-positive AD populations with and without APOE ε4 in the ADNI cohort (Table 3) are shown in Figure 5. For all the cognitive tests, the estimated longitudinal trajectories almost overlapped in the Aβ-positive AD populations with and without APOE ε4, and the AUCs of the longitudinal trajectories were not significantly different between these populations (AUC ratio: 1.20, two-sided p-value = 0.072 for MMSE, AUC ratio: 1.01, two-sided p-value = 0.618 for ADAS-Cog, and AUC ratio: 0.98, two-sided p-value = 0.122 for CDR-SOB).

**Comparison with previous findings**

*Association of AD onset with APOE ε4 status:* Hersi et al.[23] identified four systematic reviews[24–28] and five cohort studies[29–33] that assessed the association between APOE ε4 status and AD onset. Each systematic review reported the results of meta-analyses including 3–35 cohort studies (approximately 700–6,000 patients), which indicated that APOE ε4 significantly increased the risk of AD onset. Except for one cohort study[31], four cohort studies showed that APOE ε4-positive patients with MCI had a significantly increased risk of AD onset compared to APOE ε4-negative patients with MCI. The association between APOE ε4 status and AD onset for ADAS-Cog observed in our study (the time to milestone for possible AD onset was 6.5 year earlier in the Aβ-positive MCI population with APOE ε4 than in the Aβ-positive MCI population without APOE ε4 of the ADNI cohort (two-sided p-value = 0.012)) was similar to these findings.

*Association of AD progression with APOE ε4 status* Hersi et al.[23] identified the review by Allan and Ebmeier[34] as a systematic review assessing the association between APOE ε4 and AD progression. In this review, AD progression was defined based on the change in cognitive test scores from baseline to the follow-up time point of 0.5–3 years in the AD population. They identified 16 studies with 1,618 participants. The meta-analysis of these identified studies showed that the standardized mean difference of cognitive decline in cognitive test score between the AD patients with and without APOE

ε4 was 0.025 (95%CI, −0.09 to 0.14, two-sided p-value = 0.67). The proposed method evaluated the longitudinal progression of AD based on the AUC of the predicted longitudinal trajectories in each cognitive function measures, and not just the change in cognitive test scores during the short-term follow-up period. In this study, there were no significant differences in the AUCs between the Aβ-positive AD population with and without APOE ε4 for all cognitive tests. The AUC ratios of the APOE ε4-positive population relative to negative population in the ADNI cohort for MMSE, ADAS-Cog, and CDR-SOB were 1.20, 1.01, and 0.98, respectively. Therefore, the results of this study are consistent with the findings of Allan and Ebmeier[34] that instated that APOE ε4 does not accelerate the progression of AD toward further cognitive decline.

Tam et al.[3] developed a prognostic machine learning model for classifying patients with early AD (e.g., AD or MCI) into those who will show an increase in CDR-SOB until 24 months (i.e., decliners) and those who will have no change or negative changes in CDR-SOB until 24 months (i.e., stable population). This model was trained using the following features: age, sex, APOE ε4, baseline MMSE score, baseline CDR-SOB score, and volume of the anatomical brain regions extracted from MRI. By applying this model to a subset of amyloid-positive patients in the early AD population, they found that the proportion of APOE ε4 positive patients in the predicted decliners was significantly higher than that in the predictive stable population. This suggests an association between APOE ε4 status and the onset and progression of AD in early AD population. In this study, the two longitudinal trajectories of CDR-SOB in the Aβ-positive MCI and AD patients overlapped between those with and without APOE ε4 in the ADNI cohort (Figures 4C and 5C). However, according to Figure 4B (i.e., the longitudinal trajectory of ADAS-Cog in the Aβ-positive MCI population of the ADNI cohort), the time to reach a milestone for possible AD onset and the AUC showed statistically significant differences between the populations with and without APOE ε4. This suggested that APOE ε4 status may be associated with the onset and longitudinal progression of AD in the Aβ-positive MCI population..

**Application 2: Mitigating heterogeneity of disease progression among enrolled patients in Alzheimer's disease (AD) clinical trials.**

Using the proposed method, we developed an efficient AD RCT design that mitigates the heterogeneity of disease progression among enrolled patients and improves the power of statistical hypothesis testing for therapeutic effects. The proposed RCT design screens and randomizes the enrolled patients by utilizing the (real-world) data such as ADNI as follows: at the time of planning the trial, we first extracted the patient population from a real-world dataset (we used ADNI data in the following simulation study) who satisfied the eligibility criteria for the planned RCT and whose cognitive test score of the primary endpoint has been collected for at least several years (see Step 1 in the "Proposed RCT design" section in the Methods chapter along with Figure 6). Using the extracted dataset, we estimated several patterns of the longitudinal trajectories (e.g., moderate, intermediate, and rapid decline groups) on cognitive test scores. Based on our preliminary evaluation, we recommend using the Bayesian theorem for parameter estimation and their 80% credible intervals for the longitudinal trajectories (see Step 2 in the "Proposed RCT design" section). Next, before randomizing a patient to the investigational drug or placebo group, the declining slope of the cognitive test score of the primary endpoint in that patient are estimated based on the mixed-effects model used in Step I of the proposed method using the combined data of the patient's historical data and the abovementioned real-world data. If the estimated slope in the patient shows a more rapid cognitive decline than the pre-specified threshold of $W$ or no decline at all, the patient is not randomized and gets excluded from the trial (see Step 3 in the "Proposed RCT design" section). Otherwise, the patient was randomized to the drug or placebo group according to an algorithm that ensured that the drug and placebo groups were balanced with respect to the patterns of the longitudinal trajectory of the cognitive test score estimated in Step 2 of the "Proposed RCT design" section. Specifically, based on the estimated decline slope and baseline value at randomization of the cognitive test score in the enrolled patients, we calculated the predicted change in the cognitive test score from randomization to the time of primary efficacy assessment in the trial using the longitudinal trajectory of the cognitive test score (see Step 4 in the "Proposed RCT design" section). The patient is randomized to the drug or placebo group based on the stratified randomization method, with the stratum defined based on the predicted change (see Steps 5 and 6 in the "Proposed RCT design" section). The patient screening and randomization procedures are repeated until the required sample size is reached.

We conducted simulation studies to assess whether the proposed RCT design improves the operating characteristics (i.e., the accuracy of statistical inference for therapeutic effects and the power of statistical hypothesis testing) (see Figure 6). The simulation studies imitated the design and data of the phase III clinical trial of lecanemab, compared with placebo, with the primary endpoint of the change of CDR-SOB during the follow-up period of 1.5 years (Clarity AD study)[35]. Specifically, among the ADNI cohort data, we extracted the subpopulation of 230 patients with MCI and 28 patients with AD based on the following criteria so that they could imitate the patient population of the Clarity AD study: (i) baseline CDR-SOB score of 0.5 to 8.5, (ii) Aβ-positive, (iii) 50 to 90 years of age at baseline, (iv) baseline MMSE ≥ 22, and (v) follow-up of CDR-SOB ≥ 2 years and estimated the slopes $\tilde{\beta}_i$ of CDR-SOB in individuals using the proposed method (Step 1 of Figure 6). In our simulation study, we estimated the longitudinal trajectories of three decline groups ($s = 1, 2, 3$): moderate ($0 \leq \tilde{\beta}_i < 1$) ($s = 1$), intermediate ($1 \leq \tilde{\beta}_i < 2$) ($s = 2$), and rapid ($2 \leq \tilde{\beta}_i$) ($s = 3$) in the patients with MCI, and also estimated the longitudinal trajectory for the group with decline slope of $\tilde{\beta}_i \geq 0$ in the AD patients because of the small number of patients in the extracted subpopulation (Step 2 of Figure 6). For each enrolled patient, we simulated the CDR-SOB data for 1.5 years of historical period before randomization and 1.5 years of follow-up period after randomization (see the "Data generation" section in the Methods chapter). Using the simulated data for 1.5 years of historical period in the patient and the data of the extracted subpopulation, we estimated the slope of CDR-SOB in that patient (Step 3 of Figure 6) and subsequently estimated the predicted changes in CDR-SOB score during the follow-up period of 1.5 years (Step 4 of Figure 6). Utilizing the predicted changes interval (i.e., $\left(\hat{c}_i^{Lower}, \hat{c}_i^{Upper}\right)$ in Step 5 of the "Proposed RCT design" section), we employed stratified randomization (see Steps 5-6 of Figure 6). In our simulation studies, we conveniently set three randomization strata of predicted changes interval (i.e., A: 0–1.5, B: 1.5–3.0, C: 3.0–). As a representative example of using the formulae mentioned in Figure 6, when a patient with an intermediate decliner of $1 \leq \tilde{\beta}_i < 2$ and baseline CDR-SOB at randomization of 3 point is enrolled, the $\hat{c}_i^{Mean}$, $\hat{c}_i^{Lower}$, and $\hat{c}_i^{Upper}$ were 2.4, 2.1, and 2.7, respectively; therefore, this patients is included in the randomization stratum, B.

We compared the operating characteristics of the proposed RCT designs using the two prespecified thresholds of $W = 1, 2$ with those of the standard design (i.e., non-use of the proposed patient screening and randomization procedures). The standard design used the permuted block randomization method. Table 9 shows the bias and root mean square error (RMSE) of the difference estimate in the change of CDR-SOB from baseline to the time of the primary efficacy assessment (i.e., 1.5 years) between the drug and placebo groups, and the power of statistical hypothesis testing when using a mixed-effects model for repeated measures (MMRM) as the primary efficacy analysis[36], along with the average number of screened patients. The bias and RMSE of the proposed designs with $W = 1$ and 2 were smaller than those of the standard designs. The proposed design with $W = 1$ (or 2) improved the power by approximately 7% (or 3%) compared with the standard design by adding an average of 60 (or 25) patients to screen for the required number of patients in the standard design.

**Discussion**

We developed a new methodology that can predict, classify, and superimpose long-term trajectories of cognitive test scores in a self-organized manner based on the short-term follow-up data of individuals. Our methodology stands in stark contrast to current methodologies, which lack the capability to achieve such comprehensive analysis of the prediction-classification-superposition of longitudinal cognitive decline. Our internal and external validations demonstrated that APE could be controlled to within almost 10% of the maximum point of each cognitive test score, indicating the successful prediction and classification of different long-term trajectory patterns. The two practical illustrations highlight the potential effectiveness of the proposed method, thereby contributing significantly to the advancement of researches in AD.

Identification of risk factors associated with the onset and longitudinal progression of AD is important for prevention and therapeutic intervention. By combining the proposed method with the bootstrap approach, we could estimate the time to reach a milestone for possible AD onset and AUC, quantify the longitudinal progression of AD with and without risk factors of interest, and further compare the differences in these indices between the two groups using statistical hypothesis testing. According to systematic reviews and cohort studies, APOE ε4 is a strong predictor of AD onset[24]. The proposed method also found that the time to reach a milestone for possible AD onset was significantly faster in the Aβ-positive MCI population with APOE ε4 than in the Aβ-positive MCI population without APOE ε4, based on ADAS-Cog. In addition, for the association between APOE ε4 and the longitudinal progression of AD, the proposed method led to findings similar to existing evidence from systematic reviews on patients with AD[34], and a prediction study for patients with early AD[3]. Thus, the proposed method provides a new approach that enables the simultaneous identification of the timing of AD onset and risk factors for onset/longitudinal progression of AD, and is expected to contribute to disease monitoring and customized therapies for AD.

Long-term prediction of cognitive decline based on the decline speed estimated from short-term data is relevant not only for determining the timing of preventive and therapeutic interventions but also for establishing an RCT design that can mitigate the heterogeneity of disease progression among the patients enrolled in AD clinical trials. Recently, the number of RCTs for the treatment of MCI and early AD, such as aducanumab and lecanemab, has increased[35,37]. However, most clinical trials on drugs for AD have failed to detect statistically significant differences in cognitive function measures between the study period of few years between treatment and placebo groups[38]. Placebo groups in several AD clinical trials with similar inclusion and exclusion criteria have shown widely different rates of cognitive decline[39]. Moreover, AD is a heterogeneous disease with several distinct variants[13], and clinical symptoms, cognitive course, spatial patterns of brain atrophy, and tau pathology differ between the patients[40–42]. These findings suggest that the possibly failure of AD clinical trials may be attributed to the heterogeneity in rates of cognitive decline among the patients enrolled in the trials, leading to increased variation in the primary endpoint of cognitive function measures. This results in reduced statistical power of hypothesis testing of therapeutic effects. Therefore, to enhance the success probability of AD clinical trials, comprehending the disease progression trajectory of AD and controlling the heterogeneity of cognitive decline rates at the time of patient enrollment are essential. As demonstrated in Application 2, we illustrate a novel enrichment strategy for RCT that addresses the heterogeneity of cognitive decline among enrolled patients, to promote the homogeneity of disease progression between groups based on the long-term trajectory predicted by our proposed method. In simulation studies, this design decreased bias and the RMSE in the estimation of treatment effects and improved statistical power for hypothesis testing compared to the standard design. Notably, the prespecified thresholds of $W$ and randomization strata of the predicted change interval should be fine-tuned through simulation studies before trial initiation to obtain the desired operating characteristics of the proposed RCT design. Further modifications to the proposed RCT design are warranted for enhancing operating characteristics in future studies. For instance, one could consider estimating the longitudinal trajectory of cognitive test scores by including the patient historical data in Step 4 instead of Step 2 of the "Proposed RCT design" section. Additionally, Figure 3 indicates that ADAS-Cog exhibited lower fluctuations within the moderate, intermediate, and rapid cognitive decline groups than MMSE and CDR-SOB. Therefore, ADAS-Cog may be more sensitive in detecting drug effect on longitudinal cognitive decline, despite CDR-SOB being the primary endpoint in many AD clinical trials.

Many studies have delved into prediction and clustering based on biomarkers and brain atrophy, yet these studies provide predictions of relatively short-terms. Additional factors impending their widespread implementation are their high cost, invasive nature, and limited availability being restricted to a limited number of highly specialized centers. The proposed method addresses this by predicting long-term cognitive decline and quantifying disease progression heterogeneity solely through short-term cognitive test scores, without the need for blood tests or brain imaging data.

Timely identification of cognitive impairment in the elderly remains a critical healthcare priority with broad benefits for patients, their families, and the medical economy. The widespread use of digital technology in the modern world provide innovative opportunities to diagnose diseases. Beyond traditional cognitive function tests like MMSE, ADAS-Cog, and CDR-SOB, self-administered computerized cognitive devices developed for evaluating cognitive function using audiovisual tasks displayed on monitors hold much promise for clinical integration, thanks to research efforts on a global scale[43]. To summarize, this proposed method can easily be extended to predict the long-term trajectory of self-administered digitized tests. Moreover, the proposed methodology can be easily applied to predict, classify, and superimpose long-term trajectories of other neurological disorders, such as amyotrophic lateral sclerosis[44]. The implications of performing the prediction-classification-superposition of cognitive decline using the proposed method go beyond AD.

**Methods**
**Data**
Data used in this study were obtained from the ADNI and J-ADNI databases (adni.loni.usc.edu and https://humandbs.biosciencedbc.jp/en/hum0043-v1). The ADNI was launched in 2003 as a public-private partnership, led by Principal investigator Michael W. Weiner, MD. The primary goal of ADNI has been to test whether serial magnetic resonance imaging (MRI), positron emission tomography (PET), other biological markers, and clinical and neuropsychological assessments can be combined to measure the progression of MCI and early AD. For up-to-date information, see www.adni-info.org. The ADNI and J-ADNI data used in this study were downloaded on June 23, 2022 and November 16, 2020, respectively. Further, the data of J-ADNI was deposited in the National Bioscience Database Center Human Database, Japan (Research ID: hum0043.v1, 2016). The inclusion criteria for the ADNI and J-ADNI are similar to those of the MMSE (20-26), CDR-SOB (0.5-1), age (> 65 years), and NINCDS/ADRDA criteria for probable AD, and GDS (< 6)[16]. The median follow-up duration for the ADNI is 4 years, and 3 years for the J-ADNI. The intervals of cognitive examinations were 3-12 months for ADNI, and 6-12 months for J-ADNI. In the ADNI dataset, positivity for Aβ accumulation was defined as a standardized uptake value ratio (SUVR) of > 1.5 of 11C-Pittsburgh compound B (PiB) PET, SUVR > 1.11 in AV45 (florbetapir)-PET, or CSFAβ42 < 192 pg/mL[45,46]. In the J-ADNI dataset, the positivity of Aβ accumulation was defined as SUVR > 1.5 in 11C- PiB PET or a low concentration of amyloid-β 42 in the CSF (Aβ42) < 333 pg/mL[47,48]. In both datasets, we excluded participants with missing baseline data for any of the following variables: MMSE, ADAS-Cog, CDR-SOB, disease status (NC, SMC, EMCI, LMCI, or AD), age, sex, years of education, APOE ε4 allele, and Aβ positivity. We also excluded participants who were assessed only once for the MMSE, CDR-SOB, and ADAS-Cog scores after baseline assessment.

**Statistical descriptions of the four steps**

In Step I, to estimate the slope of cognitive decline (i.e., decline rate) of cognitive test scores in individuals, we applied the following mixed-effects model for measuring the change in score from the baseline value, $Y_{ij}$, of participants, $i$ ($i = 1, ..., n$) at follow-up times $t_{ij}$ ($j = 0, ..., J_i$):

$$Y_{ij} = \mathbf{x}_i^T \boldsymbol{\beta} \times t_{ij} + b_i \times t_{ij} + \varepsilon_{ij} \tag{1}$$

where, $\mathbf{x}_i^T \boldsymbol{\beta} \times t_{ij}$ is the interaction term of the disease stage of the participants at baseline assessment and at (continuous) follow-up times (years), and $\boldsymbol{\beta} = (\beta_1, \beta_2, ..., \beta_k)^T$ is the (group-specific) fixed-effects vector of the parameter (corresponding to disease stage $k$) and $\mathbf{x}_i = (x_{1i}, x_{2i}, ..., x_{ki})^T$ is the indicator vector of the disease stage for the $i$th participant. The parameter of random effect $b_i$ for participant $i$ is the individual-specific slope of cognitive decline, and $\varepsilon_{ij}$ is the random error which assumed to be followed by a normal distribution with a mean of zero and a variance of $\sigma^2$. In this model, for participant $i$ with disease stage $k$, as shown below, $\tilde{\beta}_i = \mathbf{x}_i^T \hat{\boldsymbol{\beta}} + \hat{b}_i$ is the estimated slope of the cognitive decline. The cognitive decline in AD can be influenced by various factors like genetics, lifestyle, and comorbidities. Incorporating these variables into the model of Equation (1) could provide a more comprehensive understanding of longitudinal trajectory of cognitive decline. The estimated slopes of cognitive decline in individuals $\tilde{\beta}_i$ are categorized into several groups ($s = 1, ..., S$) according to their degree of cognitive decline. The identification of distinct groups exhibiting cognitive decline can also be achieved through a data-driven approach, exemplified by the application of a finite normal mixture model to the estimated slopes of cognitive decline observed in individuals.

In Step II, the relationship between the estimated slopes $\tilde{\beta}_i$ and their individual average scores during the follow-up period $\hat{\mu}_i$ were modeled for each group using the random-effect model of $\phi(\hat{\mu}_i; \gamma_s)$. Herein, $\gamma_s$ are the subgroup-specific random effects of group $s$, and $u_i$ is the random error which assumed to be drawing from a normal distribution with a mean of zero and a variance of $\sigma_u^2$:

$$\tilde{\beta}_i = \phi(\hat{\mu}_i; \gamma_s) = \gamma_s \frac{(30 - \hat{\mu}_i)\hat{\mu}_i}{30} + u_i \text{ for MMSE}, \qquad (2)$$

$$\tilde{\beta}_i = \phi(\hat{\mu}_i; \gamma_s) = \gamma_s \frac{(\hat{\mu}_i - 85)\hat{\mu}_i}{85} + u_i \text{ for ADAS-Cog}, \qquad (3)$$

$$\tilde{\beta}_i = \phi(\hat{\mu}_i; \gamma_s) = \gamma_s \frac{(\hat{\mu}_i - 18)\hat{\mu}_i}{18} + u_i \text{ for CDR-SOB}, \qquad (4)$$

In Step III, we solved for the ordinary differential equation, $d\mu(\tau)/d\tau = \phi(\hat{\mu}_i; \hat{\gamma}_s)$, where $\tau$ is the disease progression time (year) for the purpose of predicting the temporal course (e.g., 20–30 years) in each subgroup. The initial value of the ordinary differential equation was the mean cognitive test score at baseline in the target population for the trajectory estimation. Note that the maximum (or minimum) initial values were 29, 0.5, and 1 points for MMSE, CDR-SOB, and ADAS-Cog, respectively. We performed a numerical integration to solve this ordinary differential equation between the specified interval of interest (e.g., 20–30 years) with the given initial value, obtaining the longitudinal trajectory of the cognitive test score corresponding to disease progression time $\tau$ in group, $s$ ($LTJ_s$). For participant $i$, we determine the $\tau_{i0}$ (i.e., starting point for predictive error calculation) that minimized the difference between the baseline score ($y_{i0}$) and estimated score at disease progression time $\tau$, $\hat{\mu}(\tau)$, that is, $\tau_{i0} = \arg\min|y_{i0} - \hat{\mu}(\tau)|$. The average predictive error is calculated based on, $APE = \sqrt{\frac{1}{n_s}\sum_{i=1}^{n_s}\left[\frac{1}{J_i}\sum_{j=0}^{J_i}\{y_{ij} - \hat{\mu}_s(\tau_{ij})\}^2\right]}$ using $\tau_{ij} = \tau_{i0} + t_{ij}$ for each cognitive decline group. Herein, $n_s$ is the number of participants in the cognitive decline group $s$.

**Proposed RCT design**

The procedure for patient selection and randomization of the proposed RCT design is as follows:

**Planning phase**

Step 1: We extract the patient population from a (real-world) dataset that satisfied the eligibility criteria of the planned RCT and also whose cognitive test scores of the primary endpoint in the planned RCT have been collected for at least several years.

Step 2: Using the dataset extracted in the previous step, we estimate the parameter $\gamma_s$ in the model of $\tilde{\beta}_i = \phi(\hat{\mu}_i; \gamma_s)$ based on Bayesian theorem, and obtain the posterior mean $\hat{\gamma}_s^{Mean}$, the lower limit $\hat{\gamma}_s^{Lower}$, and upper limit $\hat{\gamma}_s^{Upper}$ of the 80% Bayesian credible intervals (BCI) from posterior samples, respectively. The posterior distribution of $\gamma_s$ was computed based on 1,000 posterior samples using the noninformative prior distribution (e.g., normal distribution with a mean of zero and a variance of $10^6$) by retaining every tenth sample after discarding 2,000 iterations of burn-in in the Markov chain Monte Carlo procedure. We subsequently estimate three longitudinal trajectories: $LTJ_s^{Mean}, LTJ_s^{Lower}$, and $LTJ_s^{Upper}$ by solving the ordinary differential equations, $d\mu(\tau)/d\tau = \phi(\hat{\mu}_i; \hat{\gamma}_s^{Mean})$, $d\mu(\tau)/$ $d\tau = \phi(\hat{\mu}_i; \hat{\gamma}_s^{Lower})$ and $d\mu(\tau)/d\tau = \phi(\hat{\mu}_i; \hat{\gamma}_s^{Upper})$, respectively.

**Study phase**

Step 3: At the time of enrollment of patient $i$, the slope of the cognitive decline in that patient, $\beta_i$, is estimated by the linear mixed-effects model of Equation (1) using the combined dataset of historical data of the cognitive test score in that patient and the data extracted in Step 1. To reduce the heterogeneity of disease progression among the enrolled patients, we enroll the patients showing $0 \le \tilde{\beta}_i < W$ (e.g., $W = 1$ or $2$ for CDR-SOB).

Step 4: For an enrolled patient, $i$, the predicted changes in the cognitive test score $\hat{c}_i^{Mean}$ from baseline to the time of primary efficacy assessment and its 80% BCI ($\hat{c}_i^{Lower}, \hat{c}_i^{Upper}$) are to be calculated using the corresponding longitudinal trajectories of $LTJ_s^{Mean}, LTJ_s^{Lower}$, and $LTJ_s^{Upper}$, which are respectively estimated in Step 2. Specifically, using the baseline value of the cognitive test score in the enrolled patients, we determined the corresponding baseline disease progression times from $LTJ_s^{Mean}, LTJ_s^{Lower}$, and $LTJ_s^{Upper}$, respectively, and then determined the respective predicted cognitive test scores at the time of primary efficacy assessment (for example, after 1.5 years). The changes between the baseline and predicted values at the time of primary efficacy assessment obtained from $LTJ_s^{Mean}, LTJ_s^{Lower}$, and $LTJ_s^{Upper}$ are $\hat{c}_i^{Mean}$, $\hat{c}_i^{Lower}$, and $\hat{c}_i^{Upper}$, respectively.

Step 5: We define several intervals of the predicted change as randomization strata. If the 80% BCI of ($\hat{c}_i^{Lower}, \hat{c}_i^{Upper}$) for patient $i$ is included in one of the randomization strata, patient $i$ is randomized within this randomization stratum. Otherwise, patient $i$ is randomized within the randomization stratum of "Uncertain."

Step 6: We randomize patient $i$ to the drug or placebo group within the strata determined in the previous step.

We repeat Steps 3-6 until the required sample size of the trials is reached.

**Data generation**

In the simulation studies, for each patient, we generated the three-year CDR-SOB data (seven time points) ($t_{ij} = -1.5, -1.0, -0.5, 0, 0.5, 1.0, 1.5$). This dataset consisted of a 1.5-year historical period before randomization and 1.5-year follow-up period after randomization of the trial, followed a seven-dimensional multivariate normal distribution. To estimate the mean vector and variance-covariance matrix of the CDR-SOB data at seven time points, imitating the Clarity AD study[35], we initially extracted the CDR-SOB data from 258 patients with MCI and 76 patients with AD, based on the eligibility criteria outlined in the Clarity AD study[35] as presented in Application 2. For these populations of patients with MCI and AD, missing data at each time point for each patient are imputed using predicted values obtained from the subsequent nonlinear model utilizing the logistic function for $Z_{ij}$ of CDR-SOB for subject $i$ ($i = 1, ..., N$) at the follow-up time $t_{ij}$ ($j = 1, ..., J_i$):

$$Z_{ij} = \frac{18}{1 + \exp\{-(\alpha + a_i)(t_{ij} - \eta)\}} + v_{ij} \tag{5}$$

where $\alpha$ and $\eta$ is the fixed effects, and $a_i$ is the random effect for patient $i$ which assumed to be followed by a normal distribution with a mean of zero and a variance of $\sigma_a^2$. $(\alpha + a_i)$ means the logistic growth rate or steepness of the curve for patient $i$. $(t_{ij} - \eta)$ means a flexible horizontal translation of the logistic function. $v_{ij}$ is random error of $Z_{ij}$ which assumed to be followed by a normal distribution with a mean of zero and a variance of $\sigma_v^2$.

Consequently, by applying MMRM to the imputed dataset, we obtained the following mean vector $\widehat{m}_{MCI}$ (or $\widehat{m}_{AD}$) and variance-covariance matrix $\widehat{V}_{MCI}$ (or $\widehat{V}_{AD}$) with a heterogeneous-banded Toeplitz structure in the MCI (or AD) population:

$$\widehat{m}_{MCI} = (0.9, 1.1, 1.3, 1.6, 2.0, 2.3, 2.7),$$
$$\widehat{m}_{AD} = (2.4, 3.0, 3.9, 4.2, 4.9, 5.3, 6.0),$$

$$\widehat{V}_{MCI} = \begin{pmatrix} 1.23 & 1.27 & 1.27 & 1.11 & 1.55 & 0.85 & 1.47 \\ 1.27 & 1.77 & 1.65 & 1.38 & 2.09 & 1.16 & 1.96 \\ 1.27 & 1.65 & 2.06 & 1.60 & 2.33 & 1.41 & 2.42 \\ 1.11 & 1.38 & 1.60 & 1.67 & 2.25 & 1.31 & 2.45 \\ 1.55 & 2.09 & 2.33 & 2.25 & 4.09 & 2.21 & 3.96 \\ 0.85 & 1.16 & 1.41 & 1.31 & 2.21 & 1.60 & 2.67 \\ 1.47 & 1.96 & 2.42 & 2.45 & 3.96 & 2.67 & 5.99 \end{pmatrix}, \text{ and}$$

$$\widehat{V}_{AD} = \begin{pmatrix} 2.28 & 2.18 & 3.01 & 1.28 & 2.66 & 1.00 & 0.53 \\ 2.18 & 3.51 & 3.91 & 1.50 & 3.34 & 1.18 & 0.69 \\ 3.01 & 3.91 & 7.35 & 2.27 & 4.59 & 1.72 & 0.94 \\ 1.28 & 1.50 & 2.27 & 1.19 & 1.93 & 0.66 & 0.38 \\ 2.66 & 3.34 & 4.59 & 1.93 & 5.30 & 1.45 & 0.77 \\ 1.00 & 1.18 & 1.72 & 0.66 & 1.45 & 0.67 & 0.29 \\ 0.53 & 0.69 & 0.94 & 0.38 & 0.77 & 0.29 & 0.20 \end{pmatrix}.$$

Utilizing a 7-dimensional multivariate normal distribution, we generated CDR-SOB data at seven time points for each patient within each group. In the drug group, we introduced a drug effect by subtracting 0.2, 0.4, and 0.45 points from CDR-SOB at 0.5, 1, and 1.5 years post-randomization, respectively. Based on this premise, a sample size of 320 patients (224 for MCI and 96 for AD) was determined to achieve 84.6% power in the standard design. This calculation involved MMRM at a two-sided 5% significance level, detecting a 0.45-point difference in CDR-SOB at 1.5 years between the two groups. Employing a thousand simulated datasets, we computed the average bias and RMSE for the estimated CDR-SOB difference, along with the power of statistical hypothesis testing based on MMRM in both the proposed RCT and standard designs.

**Data and code availability**

The source data for all the tables and figures shown in this manuscript can be generated using the code provided, after downloading the datasets of ADNI and J-ADNI. All SAS code used for data processing, modeling, and figure generation can be found at: https://github.com/BioStat-85/PredictionStudy.git. For the reference, we also provide a simplified R shiny web application at https://tmdu-clinicalbiostatistics-lab.shinyapps.io/estimating-longitudinal-trajectory/ perform the estimation of the longitudinal trajectory of cognitive test score of interest using the proposed method.

**References**

1. Ansart, M. et al. Predicting the progression of mild cognitive impairment using machine learning: A systematic, quantitative and critical review. *Med. Image Anal.* **67**, 101848 (2021). 10.1016/j.media.2020.101848, Pubmed:33091740.

2. Cullen, N. C. et al. Individualized prognosis of cognitive decline and dementia in mild cognitive impairment based on plasma biomarker combinations. *Nat Aging* **1**, 114-123 (2021). 10.1038/s43587-020-00003-5, Pubmed:37117993.

3. Tam, A., Laurent, C., Gauthier, S. & Dansereau, C. Prediction of cognitive decline for enrichment of Alzheimer's disease clinical trials. *J. Prev. Alzheimers Dis.* **9**, 400-409 (2022). 10.14283/jpad.2022.49, Pubmed:35841241.

4. Kivisäkk, P. et al. Plasma biomarkers for prognosis of cognitive decline in patients with mild cognitive impairment. *Brain Commun.* **4**, fcac155 (2022). 10.1093/braincomms/fcac155, Pubmed:35800899.

5. Donohue, M. C. et al. Estimating long-term multivariate progression from short-term data. *Alzheimers Dement.* **10**(Suppl), S400-S410 (2014). 10.1016/j.jalz.2013.10.003, Pubmed:24656849.

6. Li, D., Iddi, S., Thompson, W. K., Donohue, M. C. & Alzheimer's Disease Neuroimaging Initiative. Bayesian latent time joint mixed effect models for multicohort longitudinal data. *Stat. Methods Med. Res.* **28**, 835-845 (2019). 10.1177/0962280217737566, Pubmed:29168432.

7. Raket, L. L. Statistical disease progression modeling in Alzheimer disease. *Front. Big Data* **3**, 24 (2020). 10.3389/fdata.2020.00024, Pubmed:33693397.

8. Kühnel, L., Berger, A. K., Markussen, B. & Raket, L. L. Simultaneous modeling of Alzheimer's disease progression via multiple cognitive scales. *Stat. Med.* **40**, 3251-3266 (2021). 10.1002/sim.8932, Pubmed:33853199.

9. Hirakawa, A., Sato, H., Hanazawa, R., Suzuki, K. & Japanese Alzheimer's Disease Neuroimaging Initiative. Estimating the longitudinal trajectory of cognitive function measurement using short-term data with different disease stages: application in Alzheimer's disease. *Stat. Med.* **41**, 4200-4214 (2022). 10.1002/sim.9504, Pubmed:35749990.

10. Murray, M. E. et al. Neuropathologically defined subtypes of Alzheimer's disease with distinct clinical characteristics: a retrospective study. *Lancet Neurol.* **10**, 785-796 (2011). 10.1016/S1474-4422(11)70156-9, Pubmed:21802369.

11. Habes, M. et al. Disentangling heterogeneity in Alzheimer's disease and related dementias using data-driven methods. *Biol. Psychiatry* **88**, 70-82 (2020). 10.1016/j.biopsych.2020.01.016, Pubmed:32201044.

12. Vogel, J. W. et al. Four distinct trajectories of tau deposition identified in Alzheimer's disease. *Nat. Med.* **27**, 871-881 (2021). 10.1038/s41591-021-01309-6, Pubmed:33927414.

13. Lam, B., Masellis, M., Freedman, M., Stuss, D. T. & Black, S. E. Clinical, imaging, and pathological heterogeneity of the Alzheimer's disease syndrome. *Alzheimers Res. Ther.* **5**, 1 (2013). 10.1186/alzrt155, Pubmed:23302773.

14. Tijms, B. M. et al. Pathophysiological subtypes of Alzheimer's disease based on cerebrospinal fluid proteomics. *Brain* **143**, 3776-3792 (2020). 10.1093/brain/awaa325, Pubmed:33439986.

15. Poulakis, K. et al. Multi-cohort and longitudinal Bayesian clustering study of stage and subtype in Alzheimer's disease. *Nat. Commun.* **13**, 4566 (2022). 10.1038/s41467-022-32202-6, Pubmed:35931678.

16. Cheng, D. et al. Type 2 diabetes and late-onset Alzheimer's disease. *Dement Geriatr Cogn Disord*. **31**, 424–430 (2011). 10.1159/000324134, Pubmed:21757907.

17. Rusanen, M. et al. Midlife smoking, apolipoprotein E and risk of dementia and Alzheimer's disease: a population-based cardiovascular risk factors, aging and dementia study. *Dement Geriatr Cogn Disord*. **30**: 277–284 (2010). 10.1159/000320484, Pubmed:20847559.

18. Saczynski, J.S. et al. 2010. Depressive symptoms and risk of dementia: the Framingham Heart Study. *Neurology* **75**, 35–41 (2010). 10.1212/WNL.0b013e3181e62138, Pubmed:20603483.

19. Shah, R.C. et al. Hemoglobin level in older persons and incident Alzheimer disease: prospective cohort analysis. *Neurology* **77**, 219–226 (2011). 10.1212/WNL.0b013e318225aaa9, Pubmed:21753176.

20. Helzner, EP., et al. Contribution of Vascular Risk Factors to the Progression in Alzheimer Disease. *Arch Neurol.* **66**: 343-348 (2009). 10.1001/archneur.66.3.343, Pubmed:19273753.

21. Kleiman, T. et al. Apolipoprotein E epsilon4 allele is unrelated to cognitive or functional decline in Alzheimer's disease: retrospective and prospective analysis. *Dement Geriatr Cogn Disord* **22**: 73–82 (2006). 10.1159/000093316, Pubmed:16699282.

22. Lloret, A. et al. Vitamin E paradox in Alzheimer's disease: it does not prevent loss of cognition and may even be detrimental. *J Alzheimers Dis* **17**: 143-9 (2009). 10.3233/JAD-2009-1033, Pubmed:19494439.

23. Hersi, M. et al. Risk factors associated with the onset and progression of Alzheimer's disease: A systematic review of the evidence. *NeuroToxicology* **61**, 143-187 (2017). 10.1016/j.neuro.2017.03.006, Pubmed:28363508.

24. Elias-Sonnenschein, L. S., Viechtbauer, W., Ramakers, I. H. G. B., Verhey, F. R. J. & Visser, P. J. Predictive value of APOE-4 allele for progression from MCI to AD-type dementia: a meta-analysis. *J. Neurol. Neurosurg. Psychiatry* **82**, 1149–1156 (2011). 10.1136/jnnp.2010.231555.

25. Hao, P. P., Chen, Y. G., Wang, J. L., Wang, X. L. & Zhang, Y. Meta-analysis of aldehyde dehydrogenase 2 gene polymorphism and Alzheimer's disease in East Asians. *Can. J. Neurol. Sci.* **38**, 500–506 (2011). 10.1017/s0317167100011938, Pubmed:21515512.

26. Hao, Z., Wu, B., Wang, D. & Liu, M. Association between metabolic syndrome and cognitive decline: a systematic review of prospective population-based studies. *Acta Neuropsychiatr.* **23**, 69–74 (2011). 10.1111/j.1601-5215.2011.00527.x, Pubmed:26952861.

27. Fei, M. & Jianhua, W. Apolipoprotein epsilon4-allele as a significant risk factor for conversion from mild cognitive impairment to Alzheimer's disease: a meta-analysis of prospective studies. *J. Mol. Neurosci.* **50**, 257–263 (2013). 10.1007/s12031-012-9934-y, Pubmed:23242623.

28. Liu, M., Bian, C., Zhang, J. & Wen, F. Apolipoprotein E gene polymorphism and Alzheimer's disease in Chinese population: a meta-analysis. *Sci. Rep.* **4**, 4383 (2014). 10.1038/srep04383, Pubmed:24632849.

29. Chu, L. W. et al. Bioavailable testosterone predicts a lower risk of Alzheimer's disease in older men. *J. Alzheimers Dis.* **21**, 1335–1345 (2010). 10.3233/jad-2010-100027, Pubmed:21504130.

30. Reitz, C. et al. A summary risk score for the prediction of Alzheimer disease in elderly persons. *Arch. Neurol.* **67**, 835–841 (2010). 10.1001/archneurol.2010.136, Pubmed:20625090.

31. Blasko, I. et al. Plasma amyloid beta-42 independently predicts both late-onset depression and Alzheimer disease. *Am. J. Geriatr. Psychiatry* **18**, 973–982 (2010). 10.1097/JGP.0b013e3181df48be, Pubmed:20808106.

32. Rönnemaa, E., Zethelius, B., Lannfelt, L. & Kilander, L. Vascular risk factors and dementia: 40-year follow-up of a population-based cohort. *Dement. Geriatr. Cogn. Disord.* **31**, 460–466 (2011). 10.1159/000330020, Pubmed:21791923.

33. Yang, Y. H., Roe, C. M. & Morris, J. C. Relationship between late-life hypertension, blood pressure, and Alzheimer's disease. *Am. J. Alzheimers Dis. Other Demen.* **26**, 457–462 (2011). 10.1177/1533317511421779, Pubmed:21921085.

34. Allan, C. L. & Ebmeier, K. P. The influence of ApoE4 on clinical progression of dementia: a meta-analysis. *Int. J. Geriatr. Psychiatry* **26**, 520–526 (2011). 10.1002/gps.2559, Pubmed:20845403.

35. van Dyck, C. H. et al. Lecanemab in early Alzheimer's disease. *N. Engl. J. Med.* **388**, 9-21 (2023). 10.1056/NEJMoa2212948, Pubmed:36449413.

36. Mallinckrodt, C. H., Clark, W. S. & David, S. R. Accounting for dropout bias using mixed-effects models. *J. Biopharm. Stat.* **11**, 9-21 (2001). 10.1081/BIP-100104194, Pubmed:11459446.
37. Budd Haeberlein, S. et al. Two randomized Phase 3 studies of aducanumab in early Alzheimer's disease. *J. Prev. Alzheimers Dis.* **9**, 197-210 (2022). 10.14283/jpad.2022.30, Pubmed:35542991.
38. Long, J. M. & Holtzman, D. M. Alzheimer disease: an update on pathobiology and treatment strategies. *Cell* **179**, 312–339 (2019). 10.1016/j.cell.2019.09.001, Pubmed:31564456.
39. Petersen, R. C. et al. Randomized controlled trials in mild cognitive impairment: sources of variability. *Neurology* **88**, 1751–1758 (2017). 10.1212/WNL.0000000000003907, Pubmed:28381516.
40. Scheltens, N. M. E. et al. Cognitive subtypes of probable Alzheimer's disease robustly identified in four cohorts. *Alzheimers Dement.* **13**, 1226–1236 (2017). 10.1016/j.jalz.2017.03.002, Pubmed:28427934.
41. Crane, P. K. et al. Incidence of cognitively defined late-onset Alzheimer's dementia subgroups from a prospective cohort study. *Alzheimers Dement.* **13**, 1307–1316 (2017). 10.1016/j.jalz.2017.04.011, Pubmed:28623677.
42. Ossenkoppele, R. et al. Distinct tau PET patterns in atrophy-defined subtypes of Alzheimer's disease. *Alzheimers Dement.* **16**, 335–344 (2020). 10.1016/j.jalz.2019.08.201, Pubmed:31672482.
43. Tsoy, E., Zygouris, S. & Possin, K. L. Current state of self-administered brief computerized cognitive assessments for detection of cognitive disorders in older adults: A systematic review. *J. Prev. Alzheimers Dis.* **8**, 267-276 (2021). 10.14283/jpad.2021.11, Pubmed:34101783.
44. Ramamoorthy, D. et al. Identifying patterns in amyotrophic lateral sclerosis progression from sparse longitudinal data. *Nat Comput. Sci.* **2**, 605-616 (2022). 10.1038/s43588-022-00299-w.
45. Landau, S. M. et al. Amyloid-$\beta$ imaging with Pittsburgh compound B and florbetapir: comparing radiotracers and quantification methods. *J. Nucl. Med.* **54**, 70-77 (2013). 10.2967/jnumed.112.109009, Pubmed:23166389.
46. Shaw, L. M. et al. Cerebrospinal fluid biomarker signature in Alzheimer's Disease Neuroimaging Initiative subjects. *Ann. Neurol.* **65**, 403-413 (2009). 10.1002/ana.21610, Pubmed:19296504.
47. Iwatsubo, T. et al. Japanese and North American Alzheimer's Disease Neuroimaging Initiative studies: harmonization for international trials. *Alzheimers Dement.* **14**, 1077-1087 (2018). 10.1016/j.jalz.2018.03.009, Pubmed:29753531.
48. Yamane, T. et al. Inter-rater variability of visual interpretation and comparison with quantitative evaluation of 11C-PiB PET amyloid images of the Japanese Alzheimer's Disease

Neuroimaging Initiative (J-ADNI) multicenter study. *Eur. J. Nucl. Med. Mol. Imaging* **44**, 850-857 (2017). 10.1007/s00259-016-3591-2, Pubmed:27966045.

**Acknowledgments**

This study was approved by the Ethics Committee of Tokyo Medical and Dental University (M2020-151). Data collection and sharing for this project was funded by the Alzheimer's Disease Neuroimaging Initiative (ADNI) (National Institutes of Health Grant U01 AG024904) and DOD ADNI (Department of Defense award number W81XWH-12-2-0012). ADNI is funded by the National Institute on Aging, the National Institute of Biomedical Imaging and Bioengineering, and through generous contributions from the following: AbbVie, Alzheimer's Association; Alzheimer's Drug Discovery Foundation; Araclon Biotech; BioClinica, Inc.; Biogen; Bristol-Myers Squibb Company; CereSpir, Inc.; Cogstate; Eisai Inc.; Elan Pharmaceuticals, Inc.; Eli Lilly and Company; EuroImmun; F. Hoffmann-La Roche Ltd and its affiliated company Genentech, Inc.; Fujirebio; GE Healthcare; IXICO Ltd.; Janssen Alzheimer Immunotherapy Research & Development, LLC.; Johnson & Johnson Pharmaceutical Research & Development LLC.; Lumosity; Lundbeck; Merck & Co., Inc.; Meso Scale Diagnostics, LLC.; NeuroRx Research; Neurotrack Technologies; Novartis Pharmaceuticals Corporation; Pfizer Inc.; Piramal Imaging; Servier; Takeda Pharmaceutical Company; and Transition Therapeutics. The Canadian Institutes of Health Research is providing funds to support ADNI clinical sites in Canada. Private sector contributions are facilitated by the Foundation for the National Institutes of Health (www.fnih.org). The grantee organization is the Northern California Institute for Research and Education, and the study is coordinated by the Alzheimer's Therapeutic Research Institute at the University of Southern California. ADNI data are disseminated by the Laboratory for Neuro Imaging at the University of Southern California. J-ADNI was supported by the following grants: Translational Research Promotion Project from the New Energy and Industrial Technology Development Organization of Japan; Research on Dementia, Health Labor Sciences Research Grant; Life Science Database Integration Project of Japan Science and Technology Agency; Research Association of Biotechnology (contributed by Astellas Pharma Inc., Bristol-Myers Squibb, Daiichi-Sankyo, Eisai, Eli Lilly and Company, Merck-Banyu, Mitsubishi Tanabe Pharma, Pfizer Inc., Shionogi & Co., Ltd., Sumitomo Dainippon, and Takeda Pharmaceutical Company), Japan, and a grant from an anonymous foundation.

**Competing interests**

The authors declare no potential conflict of interest.

**Figures and Tables**

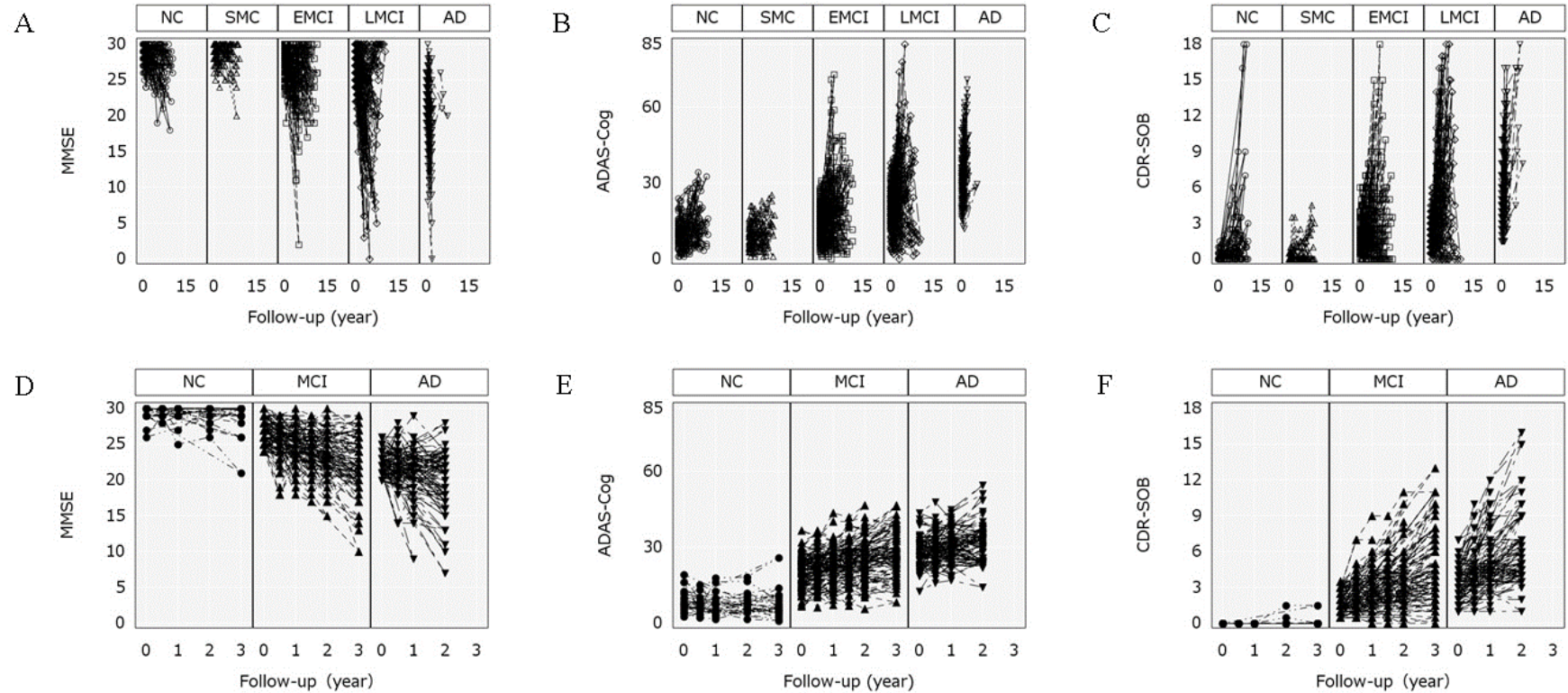

**Figure 1.** Follow-up data for each cognitive test score by baseline diagnostic stage in the Aβ-positive participants: (A) MMSE in the ADNI cohort, (B) ADAS-Cog in the ADNI cohort, (C) CDR-SOB in the ADNI cohort, (D) MMSE in the J-ADNI cohort, (E) ADAS-Cog in the J-ADNI cohort, (F) CDR-SOB in the J-ADNI cohort.
Aβ, β-amyloid peptide; ADNI, Alzheimer's Disease Neuroimaging Initiative; J-ADNI, Japanese Alzheimer's Disease Neuroimaging Initiative; NC, Normal cognitive; SMC, subjective memory complaints; EMCI, early mild cognitive impairment; MCI, mild cognitive impairment; LMCI, late mild cognitive impairment; AD, Alzheimer's Disease; MMSE, Mini-Mental State Examination; ADAS-Cog, 13-item Alzheimer's Disease Assessment Scale-Cognitive subscale; CDR-SOB, Clinical Dementia Rating-Sum of Boxes.

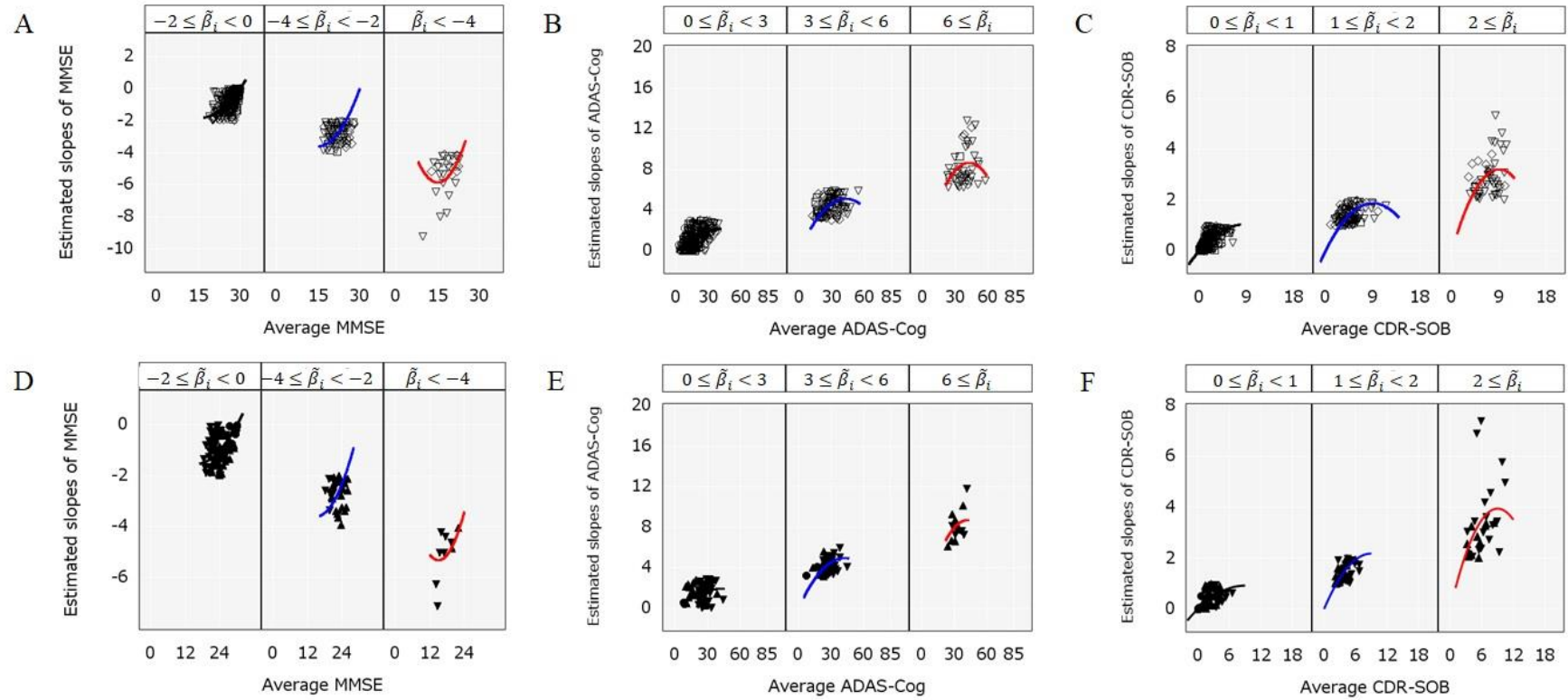

**Figure 2.** Relationship between the estimated slopes of $\tilde{\beta}_i$ and the average scores during the follow-up period of $\hat{\mu}_i$ in the Aβ-positive participants: (A) MMSE in the ADNI cohort, (B) ADAS-Cog in the ADNI cohort, (C) CDR-SOB in the ADNI cohort, (D) MMSE in the J-ADNI cohort, (E) ADAS-Cog in the J-ADNI cohort, (F) CDR-SOB in the J-ADNI cohort. The black, blue, and red solid lines represent the mean curves for the moderate, intermediate, and rapid cognitive decline groups estimated using Equation (2) to (4) of Step II, respectively.

Aβ, β-amyloid peptide; ADNI, Alzheimer's Disease Neuroimaging Initiative; J-ADNI, Japanese Alzheimer's Disease Neuroimaging Initiative; MMSE, Mini-Mental State Examination; ADAS-Cog, 13-item Alzheimer's Disease Assessment Scale-Cognitive subscale; CDR-SOB, Clinical Dementia Rating-Sum of Boxes.

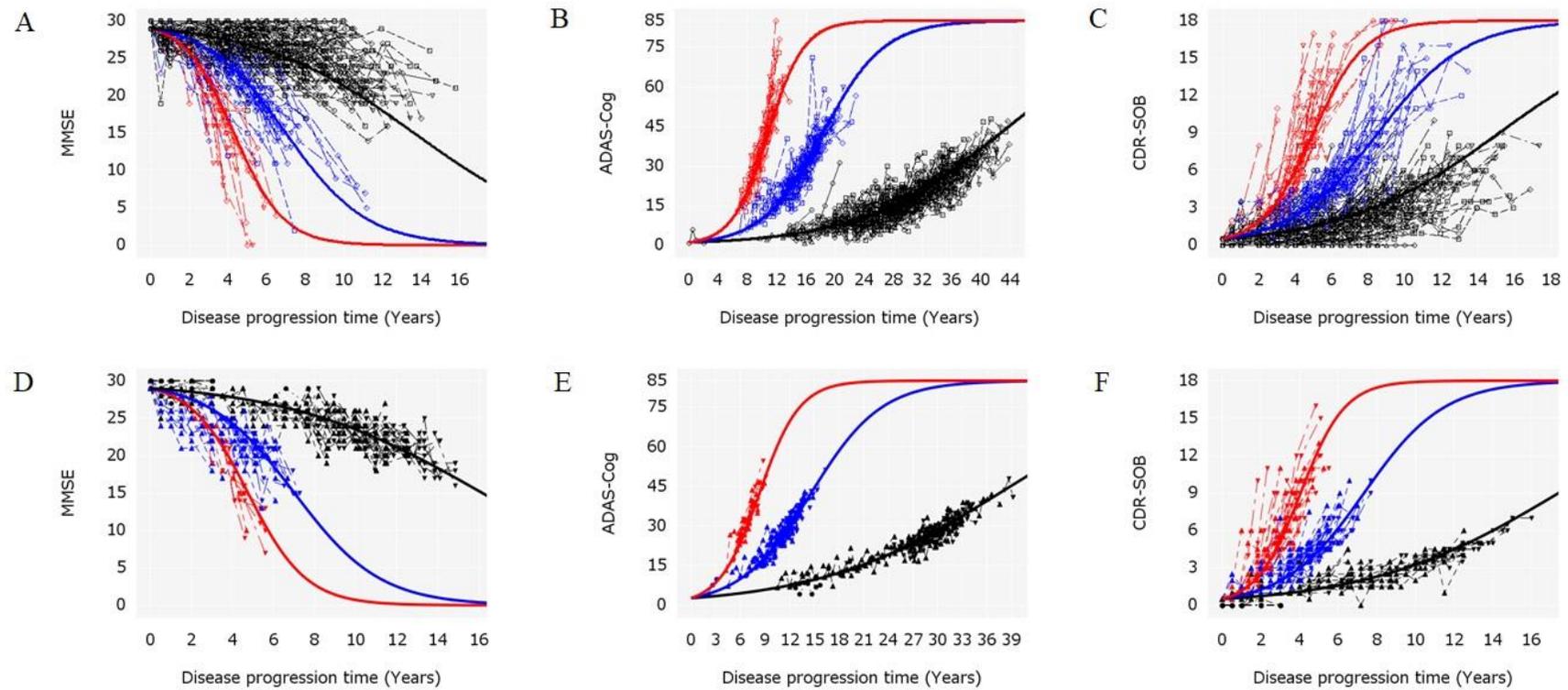

**Figure 3**. Predicted trajectories of cognitive test score for moderate, intermediate, and rapid cognitive decline groups in the Aβ-positive participants: (A) MMSE in the ADNI cohort, (B) ADAS-Cog in the ADNI cohort, (C) CDR-SOB in the ADNI cohort, (D) MMSE in the J-ADNI cohort, (E) ADAS-Cog in the J-ADNI cohort, (F) CDR-SOB in the J-ADNI cohort. The black, blue, and red solid lines represent the predicted trajectories for the moderate, intermediate, and rapid cognitive decline groups, respectively. Dash lines represent superposition of the observed actual data of cognitive test score in individuals.
Aβ, β-amyloid peptide; ADNI, Alzheimer's Disease Neuroimaging Initiative; J-ADNI, Japanese Alzheimer's Disease Neuroimaging Initiative; MMSE, Mini-Mental State Examination; ADAS-Cog, 13-item Alzheimer's Disease Assessment Scale-Cognitive subscale; CDR-SOB, Clinical Dementia Rating-Sum of Boxes.

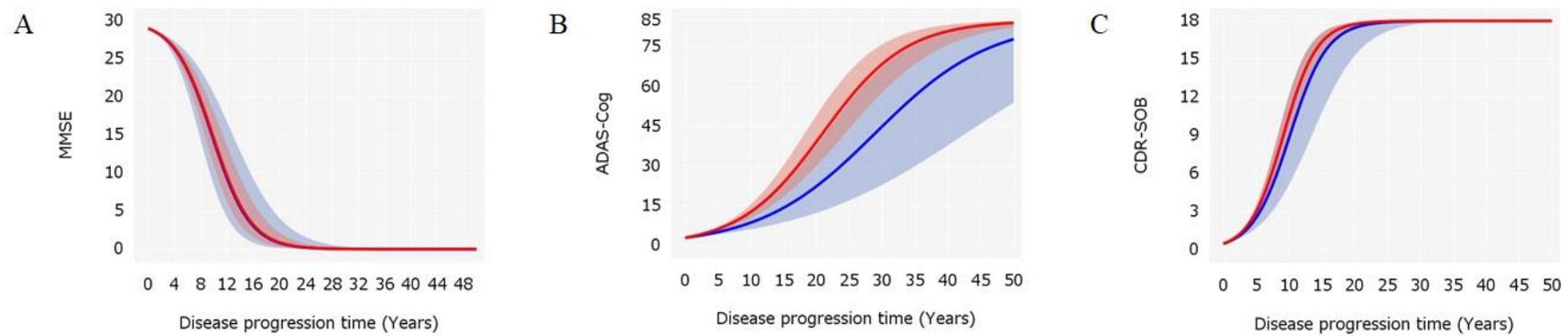

**Figure 4**. Predicted long-term trajectories of cognitive test score (A: MMSE, B: ADAS-Cog, C: CDR-SOB) by APOE ε4 status in the Aβ-positive participants with MCI in the ADNI cohort. The red and blue solid lines represent the predicted trajectories in patients with APOE ε4 positive and negative, respectively. The shaded area represents the 95% confidence interval of predicted trajectory.

MMSE, Mini-Mental State Examination; ADAS-Cog, 13-item Alzheimer's Disease Assessment Scale-Cognitive subscale; CDR-SOB, Clinical Dementia Rating-Sum of Boxes; APOE, Apolipoprotein E; Aβ, β-amyloid peptide; MCI, mild cognitive impairment; ADNI, Alzheimer's Disease Neuroimaging Initiative.

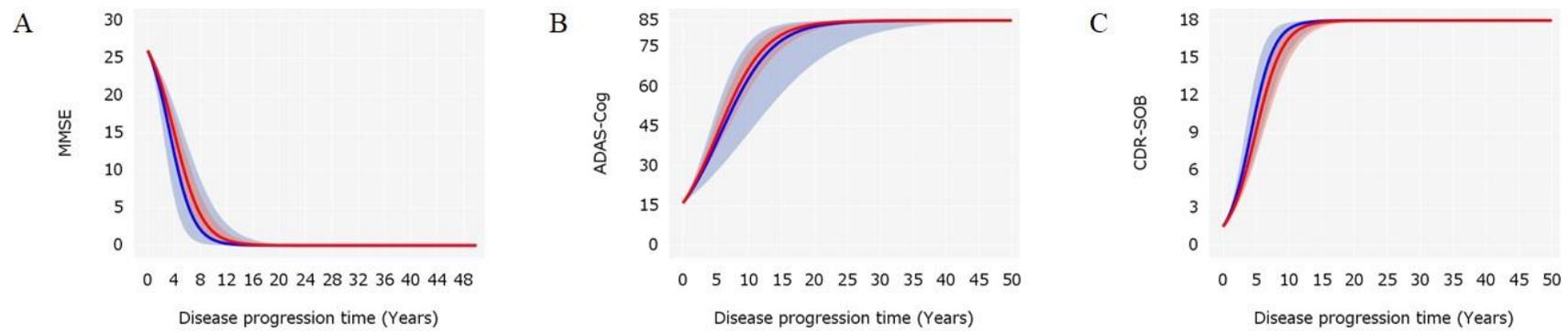

**Figure 5**. Predicted long-term trajectories of cognitive test score (A: MMSE, B: ADAS-Cog, C: CDR-SOB) by APOE ε4 status in the Aβ-positive participants with AD in the ADNI cohort. The red and blue solid lines represent the estimated trajectories for the patients with APOE ε4 positive and negative, respectively. The shaded area represents the 95% confidence interval of predicted trajectory.

MMSE, Mini-Mental State Examination; ADAS-Cog, 13-item Alzheimer's Disease Assessment Scale-Cognitive subscale; CDR-SOB, Clinical Dementia Rating-Sum of Boxes; APOE, Apolipoprotein E; Aβ, β-amyloid peptide; AD, Alzheimer's Disease; ADNI, Alzheimer's Disease Neuroimaging Initiative.

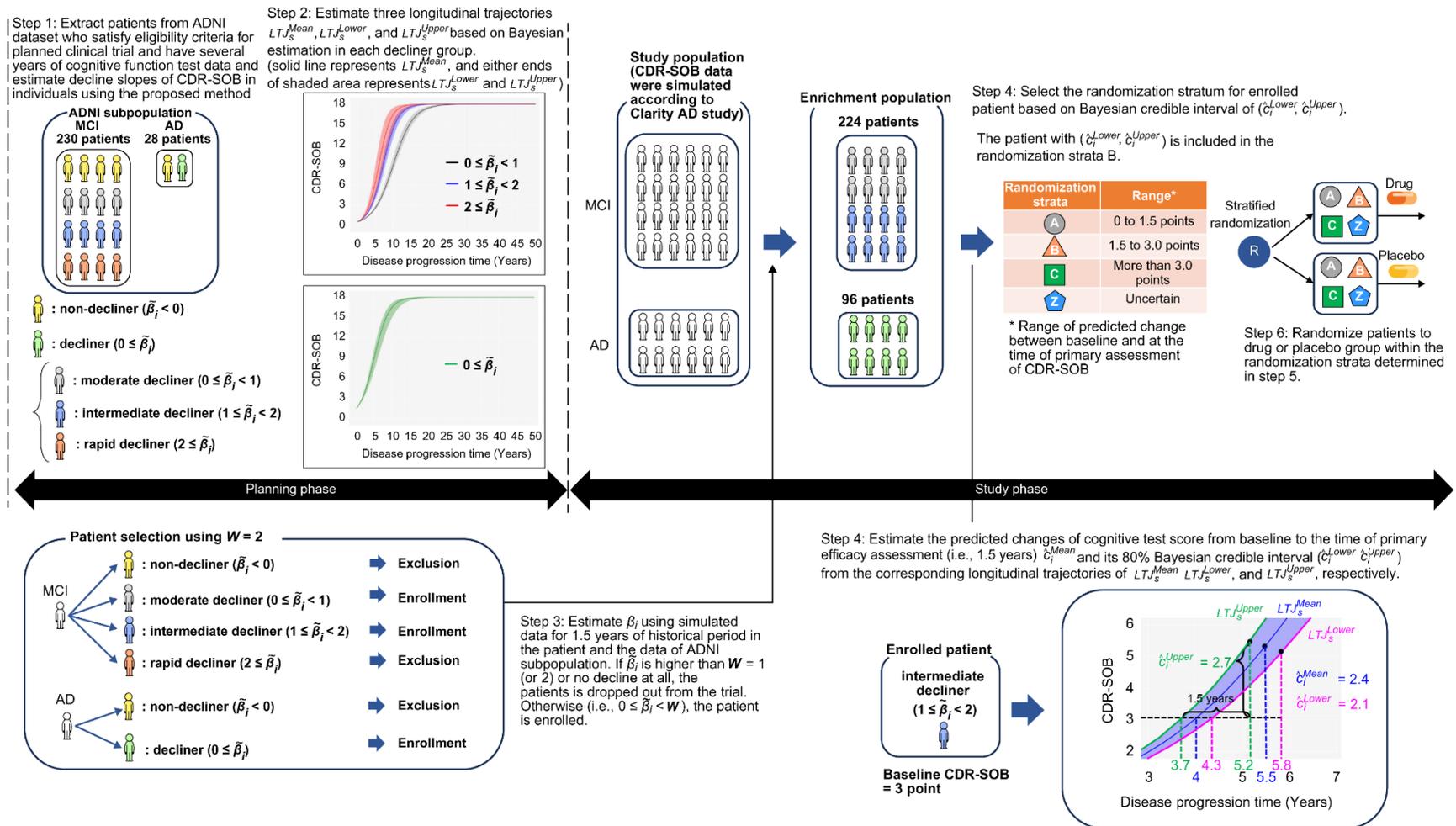

**Figure 6**. Diagram of the proposed RCT design in the simulation studies.
ADNI, Alzheimer's Disease Neuroimaging Initiative; MCI, mild cognitive impairment; AD, Alzheimer's Disease; CDR-SOB, Clinical Dementia Rating-Sum of Boxes.

**Table 1.** Participant demographics of the ADNI cohort.

|  | All | NC | SMC | EMCI | LMCI | AD |
|---|---|---|---|---|---|---|
| N | 1,503 | 336 | 176 | 337 | 385 | 269 |
| Males, N (%) | 804 (53.5%) | 156 (46.4%) | 72 (40.9%) | 188 (55.8%) | 232 (60.3%) | 156 (58%) |
| Age at baseline, median (Q1, Q3) years | 73 (68, 78) | 73 (70, 78) | 72 (67, 76) | 71 (66, 77) | 74 (69, 79) | 75 (69, 80) |
| Education, median (Q1, Q3) years | 16 (14, 18) | 16 (14, 18) | 18 (15.5, 18.5) | 16 (14, 18) | 16 (14, 18) | 16 (14, 18) |
| APOE ε4 positive, N (%) | 683 (45.4%) | 92 (27.4%) | 61 (34.7%) | 145 (43%) | 206 (53.5%) | 179 (66.5%) |
| Aβ positive, N (%) | 579 (38.5%) | 71 (21.1%) | 61 (34.7%) | 161 (47.8%) | 136 (35.3%) | 150 (55.8%) |
| MMSE score at baseline, median (Q1, Q3) | 28 (26, 29) | 29 (29, 30) | 29 (29, 30) | 29 (28, 29) | 27 (26, 29) | 23 (21, 25) |
| ADAS-Cog score at baseline, median (Q1, Q3) | 14 (9, 21.67) | 9 (6, 12) | 8 (5.33, 11.33) | 12 (9, 16) | 18.67 (14, 23.67) | 29.67 (24, 35) |
| CDR-SOB score at baseline, median (Q1, Q3) | 1 (0, 2) | 0 (0, 0) | 0 (0, 0) | 1 (0.5, 1.5) | 1.5 (1, 2) | 4.5 (3.5, 5.5) |

ADNI, Alzheimer's Disease Neuroimaging Initiative; NC, normal cognition; SMC, significant memory concern; EMCI, early mild cognitive impairment; LMCI, late mild cognitive impairment; AD, Alzheimer's Disease; N, number of participants; APOE, Apolipoprotein E; Aβ, β-amyloid peptide; MMSE, Mini-Mental State Examination; CDR-SOB, Clinical Dementia Rating-Sum of Boxes; ADAS-Cog, 13-item Alzheimer's Disease Assessment Scale-Cognitive subscale; Q1, 1st quartile; Q3, 3rd quartile.

**Table 2.** Participant demographics of the J-ADNI cohort.

|  | All | NC | MCI | AD |
|---|---|---|---|---|
| N | 288 | 84 | 116 | 88 |
| Males, N (%) | 147 (51%) | 44 (52.4%) | 62 (53.4%) | 41 (46.6%) |
| Age at baseline, year, median (Q1, Q3) | 72 (66, 77) | 67 (64, 71) | 73 (68, 77) | 75 (69, 79) |
| Education, year, median (Q1, Q3) | 13 (12, 16) | 14 (12, 16) | 13.5 (12, 16) | 12 (10, 14.5) |
| APOE ε4 positive, N (%) | 127 (44.1%) | 21 (25%) | 56 (48.3%) | 50 (56.8%) |
| Aβ positive, N (%) | 175 (60.8%) | 19 (22.6%) | 77 (66.4%) | 79 (89.8%) |
| MMSE score at baseline, median (Q1, Q3) | 26 (24, 29) | 30 (29, 30) | 27 (25, 28) | 22 (21, 24) |
| ADAS-Cog score at baseline, median (Q1, Q3) | 18.5 (10.3, 26) | 7 (4.7, 10) | 18.65 (14.3, 24.15) | 26.7 (23, 31.15) |
| CDR-SOB score at baseline, median (Q1, Q3) | 1.5 (0, 3) | 0 (0, 0) | 1.5 (1, 2) | 3.5 (2.5, 4.5) |

J-ADNI, Japanese Alzheimer's Disease Neuroimaging Initiative; NC, normal cognition; MCI, mild cognitive impairment; AD, Alzheimer's Disease; N, number of participants; APOE, Apolipoprotein E; Aβ, β-amyloid peptide; MMSE, Mini-Mental State Examination; CDR-SOB, Clinical Dementia Rating-Sum of Boxes; ADAS-Cog, 13-item Alzheimer's Disease Assessment Scale-Cognitive subscale; Q1, 1st quartile; Q3, 3rd quartile.

**Table 3.** Participant demographics of the ADNI cohort by Aβ and APOE ε4 status.

| | All | Aβ positive | Aβ negative | APOE ε4 positive | APOE ε4 negative |
|---|---|---|---|---|---|
| N | 1503 | 579 | 924 | 683 | 820 |
| Males, N (%) | 804 (53.5%) | 291 (50.3%) | 513 (55.5%) | 367 (53.7%) | 437 (53.3%) |
| Age at baseline, year, median (Q1, Q3) | 73 (68, 78) | 74 (69, 79) | 73 (67, 78) | 72 (67, 77) | 74 (69, 79) |
| Education, year, median (Q1, Q3) | 16 (14, 18) | 16 (14, 18) | 16 (14, 18) | 16 (14, 18) | 16 (14, 18) |
| APOE ε4 allele carrier, N (%) | 683 (45.4%) | 372 (64.2%) | 311 (33.7%) | 683 (100%) | 0 |
| Aβ positive, N (%) | 579 (38.5%) | 579 (100%) | 0 | 372 (54.5%) | 207 (25.2%) |
| Diagnostic at first visit in ADNI | | | | | |
|   NC, N (%) | 336 (22.4%) | 71 (12.3%) | 265 (28.7%) | 92 (13.5%) | 244 (29.8%) |
|   SMC, N (%) | 176 (11.7%) | 61 (10.5%) | 115 (12.4%) | 61 (8.9%) | 115 (14%) |
|   EMCI, N (%) | 337 (22.4%) | 161 (27.8%) | 176 (19%) | 145 (21.2%) | 192 (23.4%) |
|   LMCI, N (%) | 385 (25.6%) | 136 (23.5%) | 249 (26.9%) | 206 (30.2%) | 179 (21.8%) |
|   AD, N (%) | 269 (17.9%) | 150 (25.9%) | 119 (12.9%) | 179 (26.2%) | 90 (11%) |
| MMSE score at baseline, median (Q1, Q3) | 28 (26, 29) | 28 (25, 29) | 29 (26, 30) | 27 (25, 29) | 29 (27, 30) |
| ADAS-Cog score at baseline, median (Q1, Q3) | 14 (9, 21.67) | 17 (11, 25.67) | 12 (8, 19) | 17.7 (11, 25.67) | 12 (8, 18) |
| CDR-SOB score at baseline, median (Q1, Q3) | 1 (0, 2) | 1.5 (0.5, 3) | 0.5 (0, 2) | 1.5 (0.5, 3) | 0.5 (0, 1.5) |

ADNI, Alzheimer's Disease Neuroimaging Initiative; Aβ, β-amyloid peptide; APOE ε4, Apolipoprotein E ε4; N, number of participants; NC, normal cognition; SMC, significant memory concern; EMCI, early mild cognitive impairment; LMCI, late mild cognitive impairment; AD, Alzheimer's Disease; MMSE, Mini-Mental State Examination; CDR-SOB, Clinical Dementia Rating-Sum of Boxes; ADAS-Cog, 13-item Alzheimer's Disease Assessment Scale-Cognitive subscale; Q1, 1st quartile; Q3, 3rd quartile.

Table 4. Participant demographics of the J-ADNI cohort by Aβ and APOE ε4 status.

| | All | Aβ positive | Aβ negative | APOE ε4 positive | APOE ε4 negative |
|---|---|---|---|---|---|
| N | 288 | 175 | 113 | 127 | 161 |
| Males, N (%) | 147 (51%) | 85 (48.6%) | 62 (54.9%) | 67 (52.8%) | 80 (49.7%) |
| Age at baseline, year, median (Q1, Q3) | 72 (66, 77) | 74 (68, 77) | 69 (65, 75) | 72 (67, 76) | 71 (66, 77) |
| Education, year, median (Q1, Q3) | 13 (12, 16) | 12 (12, 16) | 14 (12, 16) | 13 (12, 16) | 13 (12, 16) |
| APOE ε4 allele carrier, N (%) | 127 (44.1%) | 109 (62.3%) | 18 (15.9%) | 127 (100%) | 0 |
| Aβ positive, N (%) | 175 (60.8%) | 175 (100%) | 0 | 109 (85.8%) | 66 (41%) |
| Diagnostic at first visit in J-ADNI | | | | | |
|   NC, N (%) | 84 (29.2%) | 19 (10.9%) | 65 (57.5%) | 21 (16.5%) | 63 (39.1%) |
|   MCI, N (%) | 116 (40.3%) | 77 (44%) | 39 (34.5%) | 56 (44.1%) | 60 (37.3%) |
|   AD, N (%) | 88 (30.6%) | 79 (45.1%) | 9 (8%) | 50 (39.4%) | 38 (23.6%) |
| MMSE score at baseline, median (Q1, Q3) | 26 (24, 29) | 25 (23, 27) | 29 (27, 30) | 26 (23, 28) | 27 (24, 30) |
| ADAS-Cog score at baseline, median (Q1, Q3) | 18.5 (10.3, 26) | 23.3 (17.3, 27.7) | 10.3 (5.7, 16.7) | 23 (16.3, 26.7) | 14.3 (7.7, 23) |
| CDR-SOB score at baseline, median (Q1, Q3) | 1.5 (0, 3) | 2 (1, 3.5) | 0.5 (0, 1.5) | 2 (1, 3.5) | 0.5 (0, 2.5) |

J-ADNI, Japanese Alzheimer's Disease Neuroimaging Initiative; Aβ, β-amyloid peptide; APOE ε4, Apolipoprotein E ε4; N, number of participants; NC, normal cognition; MCI, mild cognitive impairment; AD, Alzheimer's Disease; MMSE, Mini-Mental State Examination; CDR-SOB, Clinical Dementia Rating-Sum of Boxes; ADAS-Cog, 13-item Alzheimer's Disease Assessment Scale-Cognitive subscale; Q1, 1st quartile; Q3, 3rd quartile.

**Table 5**. Follow-up period for each cognitive function test score in the Aβ-positive participants

|  | ADNI | J-ADNI |
|---|---|---|
| MMSE |  |  |
|   N | 462 | 156 |
|   Follow-up period, year, median (Q1, Q3) | 4 (2, 5.5) | 2 (2, 3) |
| ADAS-Cog |  |  |
|   N | 451 | 156 |
|   Follow-up period, year, median (Q1, Q3) | 4 (2, 6) | 2 (2, 3) |
| CDR-SOB |  |  |
|   N | 469 | 157 |
|   Follow-up period, year, median (Q1, Q3) | 3.5 (2, 5.5) | 2 (2, 3) |

Aβ, β-amyloid peptide; ADNI, Alzheimer's Disease Neuroimaging Initiative; J-ADNI, Japanese Alzheimer's Disease Neuroimaging Initiative; MMSE, Mini-Mental State Examination; ADAS-Cog, 13-item Alzheimer's Disease Assessment Scale-Cognitive subscale; CDR-SOB, Clinical Dementia Rating-Sum of Boxes; N, number of participants; Q1, 1st quartile; Q3, 3rd quartile.

**Table 6**. Average estimated cognitive decline slope during the follow-up period in the Aβ-positive participants

| Cognitive decline groups | | ADNI | J-ADNI |
|---|---|---|---|
| MMSE | | | |
|   Moderate (Estimated slope, −2 to 0) | N | 262 | 95 |
| | Mean (SD) | −0.8 (0.6) | −0.9 (0.6) |
|   Intermediate (Estimated slope, −4 to −2) | N | 80 | 32 |
| | Mean (SD) | −2.8 (0.6) | −2.7 (0.5) |
|   Rapid (Estimated slope, < −4) | N | 26 | 9 |
| | Mean (SD) | −5.4 (1.3) | −5.1 (1.0) |
| ADAS-Cog | | | |
|   Moderate (Estimated slope, 0 to 3) | N | 219 | 64 |
| | Mean (SD) | 1.4 (0.9) | 1.6 (0.8) |
|   Intermediate (Estimated slope, 3 to 6) | N | 95 | 45 |
| | Mean (SD) | 4.5 (0.8) | 4.3 (0.7) |
|   Rapid (Estimated slope, ≥ 6) | N | 44 | 13 |
| | Mean (SD) | 8.3 (1.8) | 8.1 (1.6) |
| CDR-SOB | | | |
|   Moderate (Estimated slope, 0 to 1) | N | 262 | 73 |
| | Mean (SD) | 0.3 (0.3) | 0.4 (0.3) |
|   Intermediate (Estimated slope, 1 to 2) | N | 105 | 43 |
| | Mean (SD) | 1.4 (0.3) | 1.5 (0.3) |
|   Rapid (Estimated slope, ≥ 2) | N | 51 | 30 |
| | Mean (SD) | 2.9 (0.7) | 3.3 (1.4) |

Aβ, β-amyloid peptide; ADNI, Alzheimer's Disease Neuroimaging Initiative; J-ADNI, Japanese Alzheimer's Disease Neuroimaging Initiative; SD, standard deviation; MMSE, Mini-Mental State Examination; CDR-SOB, Clinical Dementia Rating-Sum of Boxes; ADAS-Cog, 13-item Alzheimer's Disease Assessment Scale-cognitive subscale

**Table 7**. Average predictive error in the internal and external validations of the prediction model in the Aβ-positive participants

| Cognitive decline groups | ADNI | | J-ADNI | |
|---|---|---|---|---|
| | Internal accuracy | External accuracy | Internal accuracy | External accuracy |
| MMSE | | | | |
|   Moderate (−2 to 0) | 2.12 | 1.62 | 1.66 | 1.93 |
|   Intermediate (−4 to −2) | 2.29 | 2.61 | 1.96 | 2.32 |
|   Rapid (< −4) | 3.64 | 2.48 | 3.18 | 4.00 |
| ADAS-Cog | | | | |
|   Moderate (0 to 3) | 3.82 | 2.69 | 2.65 | 3.76 |
|   Intermediate (3 to 6) | 4.27 | 2.62 | 2.64 | 4.24 |
|   Rapid (≥ 6) | 5.83 | 3.81 | 3.67 | 5.95 |
| CDR-SOB | | | | |
|   Moderate (0 to 1) | 1.06 | 0.67 | 0.65 | 0.99 |
|   Intermediate (1 to 2) | 1.23 | 1.13 | 0.99 | 1.23 |
|   Rapid (≥ 2) | 1.99 | 2.34 | 1.99 | 1.75 |

Aβ, β-amyloid peptide; ADNI, Alzheimer's Disease Neuroimaging Initiative; J-ADNI, Japanese Alzheimer's Disease Neuroimaging Initiative; MMSE, Mini-Mental State Examination; ADAS-Cog, 13-item Alzheimer's Disease Assessment Scale-Cognitive subscale; CDR-SOB, Clinical Dementia Rating-Sum of Boxes

**Table 8**. Time to reach a milestone for a possible AD onset in the ADNI cohort between two Aβ-positive populations with and without APOE ε4, its 95% CI and the statistical hypothesis testing for the difference of time to AD onset

| Cognitive test score | APOE ε4 Positive | APOE ε4 Negative | Two-sided p-value |
| --- | --- | --- | --- |
| MMSE | 6.4 [5.6, 7.4] | 6.3 [5.1, 8.4] | 0.836 |
| ADAS-Cog | 17.1 [14.9, 20.0] | 23.6 [17.5, 35.0] | 0.012 |
| CDR-SOB | 6.3 [5.8, 7.1] | 6.9 [5.8, 9.3] | 0.222 |

AD, Alzheimer's Disease; ADNI, Alzheimer's Disease Neuroimaging Initiative; Aβ, β-amyloid peptide; APOE ε4, Apolipoprotein E ε4; CI, Confidence interval; MMSE, Mini-Mental State Examination; ADAS-Cog, 13-item Alzheimer's Disease Assessment Scale-Cognitive subscale; CDR-SOB, Clinical Dementia Rating-Sum of Boxes

Note: Milestones for possible AD onset: 23 points for MMSE, 30 points for ADAS-Cog, and 4.5 point for CDR-SOB

Table 9. Bias and RMSE of difference of change of CDR-SOB from baseline to time of the primary efficacy assessment (1.5 years) between the drug and placebo groups, power of statistical hypothesis testing based on MMRM, and average number of screened patients to achieve the planned sample size of N = 320

| Method | Threshold for cognitive decline slope ($W$) | Bias | RMSE | Power | Average number of screened patients |
|---|---|---|---|---|---|
| Standard design | - | 0.008 | 0.147 | 84.6 | 320 (no use of screening) |
| Proposed design | $W = 1$ | −0.005 | 0.134 | 91.5 | 380 |
|  | $W = 2$ | −0.005 | 0.143 | 88.2 | 345 |

RMSE, Root mean square error; CDR-SOB, Clinical Dementia Rating-Sum of Boxes; MMRM, mixed-effects model for repeated measures